\documentclass[12pt]{article}
\usepackage{setspace}
\usepackage{epsfig}
\usepackage{amssymb}
\def\be{\begin{equation}}
\def\ee{\end{equation}}
\def\ba{\begin{array}}
\def\ea{\end{array}}
\def\beqn{\begin{eqnarray}}
\def\eeqn{\end{eqnarray}}
\def\nonum{\nonumber}
\def\bt{\begin{tabular}}
\def\et{\end{tabular}}
\def\bc{\begin{center}}
\def\ec{\end{center}}
\begin{document}
  \title{Texture specific mass matrices with Dirac neutrinos and their implications}

 \author{Gulsheen Ahuja$^1$, Manmohan Gupta$^1$, Monika Randhawa$^2$, Rohit Verma$^3$\\
\\
{$^1$ \it Department of Physics, Centre of Advanced Study, P.U.,
 Chandigarh, India.}\\
 {$^2$ \it University Institute of Engineering and Technology, P.U., Chandigarh, India.} \\
 {$^3$ \it Rayat Institute of Engineering and Information Technology, Ropar, India.} \\
 {\it Email: mmgupta@pu.ac.in}}

 \maketitle

\begin{abstract}
Considering Dirac neutrinos and Fritzsch-like texture 6 zero and 5
zero mass matrices, detailed predictions for cases pertaining to
normal/inverted hierarchy as well as degenerate scenario of
neutrino masses have been carried out. All the cases considered
here pertaining to inverted hierarchy and degenerate scenario of
neutrino masses are ruled out by the existing data. For the normal
hierarchy cases, the lower limit of $m_{\nu_1}$ and of $s_{13}$ as
well as the range of Dirac-like CP violating phase $\delta_l$
would have implications for the texture specific cases considered
here.
\end{abstract}

\section{Introduction}
In the last few years, apart from establishing the hypothesis of
neutrino oscillations, impressive advances have been made in
understanding the phenomenology of neutrino oscillations through
solar neutrino experiments \cite{solexp}, atmospheric neutrino
experiments \cite{atmexp}, reactor based experiments
\cite{reacexp} and accelerator based experiments \cite{accexp}.
Ever since the observation of neutrino oscillations, there has
been an explosive amount of activity both at the theoretical as
well as the experimental front in understanding the problem of
neutrino masses and mixings. In the case of neutrinos, neither the
mixing angles nor the neutrino masses show any hierarchy, this
being in sharp contrast to the distinct hierarchy shown by quark
masses and mixing angles. In fact, the two mixing angles governing
solar and atmospheric neutrino oscillations look to be rather
large, the third angle may be very small compared to these.
Further, at present there is no consensus about neutrino masses
which may show normal/inverted hierarchy or may even be
degenerate. Furthermore, the situation becomes complicated when
one realizes that neutrino masses are much smaller than the
charged fermion masses as well as it is yet not clear whether
neutrinos are Dirac or Majorana particles.

In the absence of a convincing fermion flavor theory, several
approaches have been considered \cite{9912358} to understand the
fermion mass generation problem, e.g., radiative mechanisms,
texture zeros, flavor symmetries, seesaw mechanism, extra
dimensions, etc.. In this context, texture specific mass matrices
have got good deal of attention in the literature, in particular
Fritzsch-like texture specific mass matrices seem to be very
helpful in understanding the pattern of quark mixings and CP
violation \cite{9912358, group}. Taking clue from the success of
these texture specific mass matrices in the context of quarks,
several attempts \cite{neuttex, leptex} have been made to consider
similar lepton mass matrices for explaining the pattern of
neutrino masses and mixings by using the seesaw mechanism
\cite{seesaw} given by
 \be M_{\nu}=-M_{\nu D}^T\,(M_R)^{-1}\,M_{\nu D},
\label{seesaweq} \ee \noindent where $M_{\nu D}$ and $M_R$ are
respectively the Dirac neutrino mass matrix and the right-handed
Majorana neutrino mass matrix. In order to analyze the
implications of mass matrix $M_{\nu}$, it is perhaps more
desirable to impose texture structure on $M_{\nu D}$, however in
some of the attempts \cite{leptex} texture structure has been
imposed on $M_{\nu}$ itself. It may be mentioned that although
several analyses have been carried out by considering neutrinos to
be Majorana particles, yet similar attempts have not been carried
out for Dirac neutrinos which have not yet been ruled out by
experiment \cite{diracstrumia}. In this context, several authors
have examined the possibility of Dirac neutrinos having small
masses \cite{smalldirmasses} as well as their compatibility with
the supersymmetric GUTs \cite{christoph}. This, therefore,
motivates one to consider texture specific mass matrices with
Dirac neutrinos, which are compatible with GUTs \cite{9912358,
neuttex}, as an alternative to the Majorana picture.

The possibility of having zero textures in the mass matrices of
Dirac neutrinos, with the charged leptons in the flavor basis, has
also been considered \cite{rodejohan}. In fact, in
\cite{rodejohan} a very interesting and intensive analysis has
been carried out, wherein they find that for the general form of
the Dirac neutrino mass matrix with texture 5 and 4 zeros, with
the charged leptons being in the flavor basis, one can accommodate
the current data including the possibility of one massless
neutrino, however without incorporating CP violation. Taking clues
from \cite{rodejohan} and keeping in mind broad principle like
quark-lepton symmetry \cite{qulepsym}, as well as in view of the
fact that Fritzsch-like texture specific mass matrices provide
good deal of success in understanding the quark mixing phenomenon,
we have considered similar texture specific mass matrices for the
case of Dirac neutrinos also. In this context, Fritzsch-like
texture 6 and 5 zero mass matrices provide the simplest
possibility of texture specific mass matrices. It needs to be
mentioned that the texture specific mass matrices considered in
the present work are quite different as compared to those
considered in \cite{rodejohan}, which is explained in the sequel.

In the present paper, for the case of Dirac neutrinos, we have
investigated 15 distinct possibilities of texture 6 zero and 5
zero mass matrices for normal/inverted hierarchy as well as
degenerate scenario of neutrino masses. The analysis has been
carried out by imposing Fritzsch-like texture structure on Dirac
neutrino mass matrices as well as on charged lepton mass matrices.
For the sake of completion, we have also investigated the cases
corresponding to charged leptons being in the flavor basis.
Further, detailed dependence of mixing angles on the lightest
neutrino mass as well as the parameter space available to the
phases of mass matrices have also been investigated. Furthermore,
several phenomenological quantities such as Jarlskog's rephasing
invariant parameter in the leptonic sector $J_l$ and the
corresponding CP violating Dirac-like phase $\delta_l$ have also
been calculated for different cases.

The detailed plan of the paper is as follows. In Section
(\ref{form}), we detail the essentials of the formalism connecting
the mass matrix to the neutrino mixing matrix. Inputs used in the
present analysis have been given in Section (\ref{in}). For
texture 6 zero as well as texture 5 zero mass matrices, Section
(\ref{inv}) discusses the calculations pertaining to inverted
hierarchy and degenerate scenario of neutrino masses, whereas
Section (\ref{nor}) details the analysis for normal hierarchy of
neutrino masses. Finally, Section (\ref{summ}) summarizes our
conclusions.

\section{Construction of PMNS matrix from mass matrices\label{form}}
 To begin with, we present the modified Fritzsch-like matrices, e.g.,
 \be
 M_{l}=\left( \ba{ccc}
0 & A _{l} & 0      \\ A_{l}^{*} & D_{l} &  B_{l}     \\
 0 &     B_{l}^{*}  &  C_{l} \ea \right), \qquad
M_{\nu D}=\left( \ba{ccc} 0 &A _{\nu} & 0      \\ A_{\nu}^{*} &
D_{\nu} &  B_{\nu}     \\
 0 &     B_{\nu}^{*}  &  C_{\nu} \ea \right),
 \label{frzmm5}
 \ee
$M_{l}$ and $M_{\nu D}$ respectively corresponding to Dirac-like
charged lepton and neutrino mass matrices. It may be noted that
each of the above matrix is texture 2 zero type with $A_{l(\nu)}
=|A_{l(\nu)}|e^{i\alpha_{l(\nu)}}$
 and $B_{l(\nu)} = |B_{l(\nu)}|e^{i\beta_{l(\nu)}}$, in case these
 are symmetric then $A_{l(\nu)}^*$ and $B_{l(\nu)}^*$ should be
 replaced by $A_{l(\nu)}$ and $B_{l(\nu)}$, as well as
 $C_{l(\nu)}$ and $D_{l(\nu)}$ should respectively be defined as $C_{l(\nu)}
 =|C_{l(\nu)}|e^{i\gamma_{l(\nu)}}$ and $D_{l(\nu)}
 =|D_{l(\nu)}|e^{i\omega_{l(\nu)}}$.

The texture 6 zero matrices can be obtained from the above
mentioned matrices by taking both $D_l$ and $D_{\nu}$ to be zero,
which reduces the matrices $M_{l}$ and $M_{\nu D}$ each to texture
3 zero type. Texture 5 zero matrices can be obtained by taking
either $D_l=0$ and $D_{\nu}\neq 0$ or $D_{\nu}=0$ and $D_l \neq
0$, thereby, giving rise to two possible cases of texture 5 zero
matrices, referred to as texture 5 zero $D_l=0$ case pertaining to
$M_l$ texture 3 zero type and $M_{\nu D}$ texture 2 zero type and
texture 5 zero $D_{\nu}=0$ case pertaining to $M_l$ texture 2 zero
type and $M_{\nu D}$ texture 3 zero type. It may be added that the
above formulation of texture zeros of mass matrices is quite
different as compared to that considered in \cite{rodejohan},
wherein the authors have examined the minimal allowed structure of
the Dirac neutrino mass matrix, referred to as $m_{\nu}$ by them
and $M_{\nu D}$ by us. They have carried out the analysis by
considering charged lepton mass matrix to be diagonal and
incorporating upto 5 zero entries in $m_{\nu}$ alone, unlike the
way we have defined texture zero mass matrices above, essentially
involving both $M_l$ and $M_{\nu D}$.

To fix the notations and conventions as well as to facilitate the
understanding of inverted hierarchy case and its relationship to
the normal hierarchy case, we detail the formalism connecting the
mass matrix to the neutrino mixing matrix. The mass matrices $M_l$
and $M_{\nu D}$ given in equation (\ref{frzmm5}), for hermitian as
well as symmetric case, can be exactly diagonalized. Details of
hermitian case can be looked up in our earlier work \cite{group},
the symmetric case can similarly be worked out. To facilitate
diagonalization, the mass matrix $M_k$, where $k=l, \nu D$, can be
expressed as
\be
M_k= Q_k M_k^r P_k \,  \label{mk} \ee or  \be M_k^r= Q_k^{\dagger}
M_k P_k^{\dagger}\,, \label{mkr} \ee where $M_k^r$ is a real
symmetric matrix with real eigenvalues and $Q_k$ and $P_k$ are
diagonal phase matrices. For the hermitian case $Q_k=
P_k^{\dagger}$, whereas for the symmetric case under certain
conditions $Q_k= P_k$. In general, the real matrix $M_k^r$ is
diagonalized by the orthogonal transformation $O_k$, e.g., \be
M_k^{diag}= {O_k}^T M_k^r O_k \,, \label{mkdiag} \ee which on
using equation (\ref{mkr}) can be rewritten as \be M_k^{diag}=
{O_k}^T Q_k^{\dagger} M_k P_k^{\dagger} O_k \,. \label{mkdiag2}
\ee To facilitate the construction of diagonalization
transformations for different hierarchies, we introduce a diagonal
phase matrix $\xi_k$ defined as $ {\rm diag} (1,\,e^{i \pi},\,1)$
for the case of normal hierarchy and as $ {\rm diag} (1,\,e^{i
\pi},\,e^{i \pi})$ for the case of inverted hierarchy. Equation
(\ref{mkdiag2}) can now be written as \be \xi_k M_k^{diag}=
{O_k}^T Q_k^{\dagger} M_k P_k^{\dagger} O_k \,, \label{mkdiag3}
\ee which can also be expressed as \be M_k^{diag}= \xi_k^{\dagger}
{O_k}^T Q_k^{\dagger} M_k P_k^{\dagger} O_k \,. \label{mkdiag4}
\ee Making use of the fact that $O_k^*=O_k$ it can be further
expressed as
\be
M_k^{diag}=(Q_k O_k \xi_k)^{\dagger} M_k (P_k^{\dagger}
O_k),\label{mkeq} \ee from which one gets \be M_k=Q_k O_k \xi_k
M_k^{diag} O_k^T P_k.\label{mkeq2} \ee

The case of leptons is fairly straight forward, for the Dirac
neutrinos the diagonalizing transformation is hierarchy specific.
To clarify this point further, in analogy with equation
(\ref{mkeq2}), we can express $M_{\nu D}$ as \be M_{\nu D}=Q_{\nu
D} O_{\nu D} \xi_{\nu D} M_{\nu D}^{diag} O_{\nu D}^T P_{\nu
D}.\label{mnud} \ee

The lepton mixing matrix, obtained from the matrices used for
diagonalizing the mass matrices $M_l$ and $M_{\nu D}$, is
expressed as
 \be
U =(Q_l O_l \xi_l)^{\dagger} (P_{\nu D} O_{\nu D}\xi_{\nu D}).
\label{mix5} \ee Eliminating the phase matrices $\xi_l$ and
$\xi_{\nu D}$ by redefinition of the charged lepton and Dirac
neutrinos, the above equation becomes
\be
 U = O_l^{\dagger} Q_l P_{\nu D} O_{\nu D} \,, \label{mixreal} \ee
where $Q_l P_{\nu D}$, without loss of generality, can be taken as
$(e^{i\phi_1},\,1,\,e^{i\phi_2})$, $\phi_1$ and $\phi_2$ being
related to the phases of mass matrices and can be treated as free
parameters.

To understand the relationship between diagonalizing
transformations for different hierarchies of neutrino masses as
well as their relationship with the charged lepton case, we
reproduce the general diagonalizing transformation $O_k$. The
elements of $O_k$ can figure with different phase possibilities,
however these possibilities are related to each other through
phase matrices $Q_l$ and $P_{\nu D}.$ For the present work, we
have chosen the possibility, \be O_k= \left( \ba{ccc} ~~O_k(11)&
~~O_k(12)& ~O_k(13)
\\
 ~~O_k(21)& -O_k(22)& ~O_k(23)\\
     -O_k(31) & ~~O_k(32) & ~O_k(33) \ea \right), \ee
where \beqn O_k(11) & = & {\sqrt \frac{m_{2} m_{3}
(m_{3}-m_{2}-D_k)}
     {(m_{1}-m_{2}+m_{3}-D_k)
(m_{3}-m_{1})(m_{1}+m_{2})} } \nonum  \\ O_k(12) & = & {\sqrt
\frac{m_{1} m_{3}
 (m_{1}+m_{3}-D_k)}
   {(m_{1}-m_{2}+m_{3}-D_k)
 (m_{2}+m_{3})(m_{1}+m_{2})} }
\nonum   \\O_k(13) & = & {\sqrt \frac{m_{1} m_{2}
 (m_{2}-m_{1}+D_k)}
    {(m_{1}-m_{2}+m_{3}-D_k)
(m_{2}+m_{3})(m_{3}-m_{1})} } \nonum   \\ O_k(21) & = & {\sqrt
\frac{m_{1}
 (m_{3}-m_{2}-D_k)}
  {(m_{3}-m_{1})(m_{1}+m_{2})} }
\nonum  \\O_k(22) & = & {\sqrt \frac{m_{2} (m_{1}+m_{3}-D_k)}
  {(m_{2}+m_{3})(m_{1}+m_{2})} }
 \nonum    \\
O_k(23) & = & \sqrt{\frac{m_3(m_{2}-m_{1}+D_k)}
 {(m_{2}+m_{3})(m_{3}-m_{1})} }
\nonum   \\O_k(31) & = &
 \sqrt{\frac{m_{1} (m_{2}-m_{1}+D_k)
    (m_{1}+m_{3}-D_k)}
{(m_{1}-m_{2}+m_{3}-D_k)(m_{1}+m_{2})(m_{3}-m_{1})}} \nonum
\\O_k(32) & = & {\sqrt \frac{m_{2}(m_{2}-m_{1}+D_k)
(m_{3}-m_{2}-D_k)}{(m_{1}-m_{2}+m_{3}-D_k)
 (m_{2}+m_{3})(m_{1}+m_{2})} }
 \nonum  \\
O_k(33) & = & {\sqrt \frac{m_{3}(m_{3}-m_{2}-D_k)
(m_{1}+m_{3}-D_k)}{(m_{1}-m_{2}+m_{3}-D_k)
 (m_{3}-m_{1})(m_{2}+m_{3})}} \label{diageq} \,,
 \eeqn  $m_1$, $-m_2$,
$m_3$ being the eigenvalues of $M_k$. In the case of charged
leptons, because of the hierarchy $m_e \ll m_{\mu} \ll m_{\tau}$,
the mass
 eigenstates can be approximated respectively to the flavor
eigenstates as has been considered by several authors \cite{
xingn5, fuku5}. Using the approximation, $m_{l1} \simeq m_e$,
$m_{l2} \simeq m_{\mu}$ and $m_{l3} \simeq m_{\tau}$, the first
element of the matrix $O_l$ can be obtained from the corresponding
element of equation (\ref{diageq}) by replacing $m_1$, $-m_2$,
$m_3$ with $m_e$, $-m_{\mu}$, $m_{\tau}$, e.g.,
 \be  O_l(11) = {\sqrt
\frac{m_{\mu} m_{\tau} (m_{\tau}-m_{\mu}-D_l)}
     {(m_{e}-m_{\mu}+m_{\tau}-D_l)
(m_{\tau}-m_{e})(m_{e}+m_{\mu})} } ~. \ee

For normal hierarchy defined as $m_{\nu_1}<m_{\nu_2}\ll
m_{\nu_3}$, as well as for the corresponding degenerate case given
by $m_{\nu_1} \lesssim m_{\nu_2} \sim m_{\nu_3}$, equation
(\ref{diageq}) can also be used to obtain the first element of
diagonalizing transformation for Dirac neutrinos. This element can
be obtained from the corresponding element of equation
(\ref{diageq}) by replacing $m_1$, $-m_2$, $m_3$ with $m_{\nu 1}$,
$-m_{\nu 2}$, $m_{\nu 3}$ and is given by
 \be O_{\nu D}(11)  =  {\sqrt \frac{m_{\nu_2} m_{\nu 3} (m_{\nu 3}-m_{\nu 2}-D_{\nu})}
     {(m_{\nu 1}-m_{\nu 2}+m_{\nu 3}-D_{\nu})
(m_{\nu 3}-m_{\nu 1})(m_{\nu 1}+m_{\nu 2})} }, \ee where
$m_{\nu_1}$, $m_{\nu_2}$ and $m_{\nu_3}$ are neutrino masses.

In the same manner, one can obtain the elements of diagonalizing
transformation for the inverted hierarchy case defined as
$m_{\nu_3} \ll m_{\nu_1} < m_{\nu_2}$ as well as for the
corresponding degenerate case given by $m_{\nu_3} \sim m_{\nu_1}
\lesssim m_{\nu_2}$. The corresponding first element, obtained by
replacing $m_1$, $-m_2$, $m_3$ with $m_{\nu 1}$, $-m_{\nu 2}$,
$-m_{\nu 3}$ in equation (\ref{diageq}), is given by
 \be O_{\nu D}(11)  =  {\sqrt \frac{m_{\nu_2} m_{\nu 3} (m_{\nu 3}+m_{\nu 2}+D_{\nu})}
     {(-m_{\nu 1}+m_{\nu 2}+m_{\nu 3}+D_{\nu})
(m_{\nu 3}+m_{\nu 1})(m_{\nu 1}+m_{\nu 2})} }. \ee The other
elements of diagonalizing transformations in the case of neutrinos
as well as charged leptons can similarly be found. Detailed
expressions of the Pontecorvo-Maki-Nakagawa-Sakata (PMNS) matrix
elements \cite{pmns} have been presented in Appendix.

\section{Inputs used in the present analysis\label{in}}
Before going into the details of the analysis, we would like to
mention some of the essentials pertaining to various inputs.
Adopting the three neutrino framework, several authors
\cite{valle}-\cite{ fogli} have presented updated information
regarding the neutrino mass and mixing parameters obtained by
carrying out detailed global analyses. The latest situation
regarding masses and mixing angles at 3$\sigma$ C.L. is summarized
as follows \cite{fogli},
\be
 \Delta m_{12}^{2} = (7.14 - 8.19)\times
 10^{-5}~\rm{eV}^{2},~~~~
 \Delta {\it m}_{23}^{2} = (2.06 - 2.81)\times 10^{-3}~ \rm{eV}^{2},
 \label{solatmmass}\ee
\be
{\rm sin}^2\,\theta_{12}  =  0.263 - 0.375,~~~
 {\rm sin}^2\,\theta_{23}  =  0.331 - 0.644,~~~
 {\rm sin}^2\,\theta_{13} \leq 0.046. \label{s13}
\ee The above data reveals that at present not much is known about
the hierarchy of neutrino masses as well as about their absolute
values.

The masses and mixing angles, used in the analysis, have been
constrained by the data given in equations (\ref{solatmmass}) and
(\ref{s13}). For the purpose of calculations, we have taken the
lightest neutrino mass, the phases $\phi_1$, $\phi_2$ and $D_{l,
\nu}$ as free parameters, the other two masses are constrained by
$\Delta m_{12}^2 = m_{\nu_2}^2 - m_{\nu_1}^2 $ and $\Delta
m_{23}^2 = m_{\nu_3}^2 - m_{\nu_2}^2 $ in the normal hierarchy
case and by $\Delta m_{23}^2 = m_{\nu_2}^2 - m_{\nu_3}^2$ in the
inverted hierarchy case. It may be noted that lightest neutrino
mass corresponds to $m_{\nu_1}$ for the normal hierarchy case and
to $m_{\nu_3}$ for the inverted hierarchy case. In the case of
normal hierarchy, the explored range for $m_{\nu_1}$ is taken to
be $0.0001\,\rm{eV}-1.0\,\rm{eV}$, which is essentially governed
by the mixing angle $s_{12}$, related to the ratio
$\frac{m_{\nu_1}}{m_{\nu_2}}$. For the inverted hierarchy case
also we have taken the same range for $m_{\nu_3}$ as our
conclusions remain unaffected even if the range is extended
further. In the absence of any constraint on the phases, $\phi_1$
and $\phi_2$ have been given full variation from 0 to $2\pi$.
Although $D_{l, \nu}$ are free parameters, however, they have been
constrained such that diagonalizing transformations, $O_l$ and
$O_{\nu}$, always remain real, implying $D_{l}< m_{l_3} - m_{l_2}$
whereas $D_{\nu} < m_{\nu_3} - m_{\nu_2}$ for normal hierarchy and
$D_{\nu} < m_{\nu_1} - m_{\nu_3}$ for inverted hierarchy.

We have carried out detailed calculations pertaining to texture 6
zero as well as two possible cases of texture 5 zero lepton mass
matrices, e.g., $D_l=0$ case and $D_{\nu}=0$ case. Corresponding
to each of these cases, we have considered three possibilities of
neutrino masses having normal/inverted hierarchy or being
degenerate. In addition to these 9 possibilities, we have also
considered those cases when the charged leptons are in the flavor
basis. These possibilities sum up to 18, however, the texture 5
zero $D_{\nu}=0$ case with charged leptons in the flavor basis
reduces to the similar texture 6 zero case, hence the 18
possibilities reduce to 15 distinct cases.

\section{Inverted hierarchy and degenerate scenario of neutrino masses\label{inv}}
\subsection{Texture 6 zero mass matrices}
To begin with, we first consider the cases pertaining to inverted
hierarchy of neutrino masses as well as when neutrino masses are
degenerate. Interestingly, we find that all the cases pertaining
to inverted hierarchy and degenerate scenarios of neutrino masses
seem to be ruled out. This can be concluded using the plots of the
variation of the mixing angles with the lightest neutrino mass.
For the texture 6 zero case, in Figure (\ref{nhih6zdir}), by
giving full variations to other parameters, we have plotted the
mixing angles against the lightest neutrino mass. The dotted lines
and the dot-dashed lines depict the limits obtained assuming
normal and inverted hierarchy respectively, the solid horizontal
lines show the 3$\sigma$ limits of the plotted mixing angle as
given in equation (\ref{s13}). A look at Figure (\ref{nhih6zdir}a)
shows that for $m_{\nu_1} \sim 0.0001~\rm{eV}$ there is a slight
overlap of the inverted hierarchy region with the experimental
limits of the angle $s_{12}$. Similarly, Figure (\ref{nhih6zdir}b)
shows that again for $m_{\nu_1} \sim 0.1-1~\rm{eV}$ there is an
overlap of the inverted hierarchy region with the experimental
limits of the angle $s_{13}$. However, it is easily evident from
Figure (\ref{nhih6zdir}c) that inverted hierarchy is ruled out at
3$\sigma$ C.L. by the experimental limits on the mixing angle
$s_{23}$. It may be emphasized that in Figures (\ref{nhih6zdir}a)
and (\ref{nhih6zdir}b) a slight overlap of the inverted hierarchy
region with the experimental limits of the two angles does not
affect our conclusions regarding inverted hierarchy of neutrino
masses since to rule it out it is sufficient to do so from any one
of the graphs.

One can easily check that degenerate scenarios characterized by
either $m_{\nu_1} \lesssim m_{\nu_2} \sim m_{\nu_3} \sim
0.1~\rm{eV}$ or $m_{\nu_3} \sim m_{\nu_1} \lesssim m_{\nu_2} \sim
0.1~\rm{eV}$ are clearly ruled out from Figures (\ref{nhih6zdir}a)
and (\ref{nhih6zdir}c). This can be understood by noting that
around $0.1~\rm{eV}$, the limits obtained assuming normal and
inverted hierarchies have no overlap with the experimental limits
of angles $s_{12}$ and $s_{23}$.

\subsection{Texture 5 zero mass matrices}
Coming to the texture 5 zero cases, we first discuss the case when
$D_l=0$ and $D_{\nu} \neq 0$. In Figure (\ref{nhih5zdndir}) we
have plotted the mixing angles against the lightest neutrino mass
for both normal and inverted hierarchy for a particular value of
$D_{\nu}= \sqrt{m_{\nu_3}}$. A look at figures
(\ref{nhih5zdndir}a) and (\ref{nhih5zdndir}c) reveals that the
region pertaining to inverted hierarchy, depicted by dot-dashed
lines, shows an overlap with the experimental limits on $s_{12}$
and $s_{23}$ respectively. The graph of $s_{13}$ versus the
lightest neutrino mass, shown in Figure (\ref{nhih5zdndir}b)
immediately rules out inverted hierarchy by experimental limits on
angle $s_{13}$.

In Figure (\ref{dmlimitsihdir}) we have plotted allowed parameter
space for the three mixing angles in the $D_{\nu}-$lightest
neutrino mass plane, for texture 5 zero $D_l=0$ case. This allows
us to extend our results to other acceptable values of $D_{\nu}$
and study their implications. Figure (\ref{dmlimitsihdir}) reveals
that the allowed parameter spaces of the three mixing angles show
an overlap only when $D_{\nu}\sim0$, which leads to the present
texture 6 zero case, wherein degenerate scenario has already been
ruled out. Therefore, again one can easily conclude that inverted
hierarchy as well as degenerate scenarios are ruled out for
texture 5 zero $D_l=0$ case, not only for
$D_{\nu}=\sqrt{m_{\nu_3}}$ but also for its other allowed values.

For the texture 5 zero $D_{\nu}=0$ and $D_l \neq 0$ case, the
plots of mixing angles against the lightest neutrino mass are
shown in Figure (\ref{nhih5zdldir}). Interestingly, these graphs
are very similar to Figure (\ref{nhih6zdir}) pertaining to the
texture 6 zero case of Dirac neutrinos. Therefore, arguments
similar to the ones for the texture 6 zero case lead us to
conclude that both inverted hierarchy as well as degenerate
scenarios of neutrino masses are ruled out for this case as well.
It may be mentioned that similarities observed in the mixing
angles variation with the lightest neutrino mass for the texture 5
zero $D_{\nu}=0$ and the texture 6 zero case can be understood by
noting that a very strong hierarchy in the case of charged leptons
reduces the texture 5 zero $D_{\nu}=0$ case essentially to the
texture 6 zero case only.

In case charged lepton mass matrices are considered to be in the
flavor basis, one can easily find, using the above methodology,
that all the cases pertaining to inverted hierarchy and degenerate
scenario of neutrino masses are again ruled out.

\section{Normal hierarchy of neutrino masses\label{nor}}
\subsection{Texture 6 zero mass matrices}
After considering the implications of the texture 6 zero as well
as the two cases of texture 5 zero mass matrices on inverted
hierarchy of neutrino masses as well as neutrino masses being
degenerate, we come to case of normal hierarchy of neutrino
masses. In Table (\ref{tab1dir}) we have presented the viable
ranges of neutrino masses, mixing angle $s_{13}$, Jarlskog's
rephasing invariant parameter in the leptonic sector $J_l$ and the
Dirac-like CP violating phase in the leptonic sector $\delta_l$.
In the texture 6 zero case, the possibility of charged leptons
being in the flavor basis is completely ruled out, therefore in
the table we have presented the results corresponding to the case
when $M_l$ is considered texture specific. A general look at the
table reveals several interesting points. In particular, for the
texture 6 zero matrices, the viable ranges of masses $m_{\nu_1}$,
$m_{\nu_2}$ and $m_{\nu_3}$ for Dirac neutrinos are quite narrow.
Also, one can easily see that the range of the angle $s_{13}$ is
again quite narrow, particularly its upper limit being quite
small. Therefore, a measurement of $s_{13}$ would have direct
implications for this case. Also, the range of Jarlskog's
rephasing invariant parameter $J_l$ is quite narrow for this case
in comparison with its expectation from the mixing matrix
\cite{Jllimit}. For this case of texture 6 zero matrices, we have
also examined the implications of the mixing angle $s_{13}$ on the
phases $\phi_1$ and $\phi_2$. In this context, in Figure
(\ref{s13cont6zdir}) we have plotted the contours for $s_{13}$ in
$\phi_1 - \phi_2$ plane. These contours indicate that the mixing
angle $s_{13}$ constrains both the phases $\phi_1$ and $\phi_2$.
For example, if the lower limit of $s_{13}$ around 0.07, then
$\phi_1$ lies in either the I or the IV quadrant and $\phi_2$ lies
between 135$^{\circ}$ - 225 $^{\circ}$.

\subsection{Texture 5 zero mass matrices}
Coming to the texture 5 zero mass matrices, we would like to
mention that out of the two possible cases, for the texture 5 zero
$D_{\nu}=0$ case, again the possibility of $M_l$ being in the
flavor basis does not yield any results. However, for the $D_l=0$
case, both the possibilities of $M_l$ having Fritzsch-like
structure as well as $M_l$ being in the flavor basis yield viable
ranges for the various phenomenological quantities. In Table
(\ref{tab1dir}), we have presented the results corresponding to
$M_l$ having Fritzsch-like structure. A comparison of the texture
5 zero $D_l=0$ case with the above mentioned texture 6 zero
matrices reveals several interesting points. From the table one
finds that going from texture 6 zero to texture 5 zero $D_l=0$
case, the viable range of $m_{\nu_1}$ gets much broader, in the
texture 6 zero case it being almost a unique value. Similarly,
from almost a unique value of $m_{\nu_3}$ for the texture 6 zero
matrices, now one gets a range of $m_{\nu_3}$. Also, it may be
seen that the upper limit of $s_{13}$ is pushed considerably
higher which can be understood by noting that $s_{13}$ is quite
sensitive to variations in $D_{\nu}$. Similarly, as expected the
ranges of $J_l$ and $\delta_l$ become broader as compared to the
texture 6 zero case. The Pontecorvo-Maki-Nakagawa-Sakata (PMNS)
mixing matrix \cite{pmns} obtained for the texture 5 zero $D_l=0$
case is given by
 \be U=\left( \ba{ccc}
 0.7897  -  0.8600   &    0.5036  -  0.5998 &      0.0054  -  0.1600 \\
  0.1838  -  0.4748   &    0.4859  -  0.7438 &      0.5726  -  0.8194 \\
  0.3107  -  0.5633   &    0.3974  -  0.6890 &      0.5650  -
  0.8142
 \ea \right). \label{dirmat}\ee

For the case of $M_l$ being in the flavor basis, the range of
masses so obtained are $m_{\nu_1}=0.0020-0.0040$,
$m_{\nu_2}=0.0088-0.0100$ and $m_{\nu_3}=0.0422-0.0548$. The range
of the mixing angle $s_{13}$ is $0.0892-0.1594$, indicating that
the lower limit of $s_{13}$ is considerably high which implies
that refinements in the measurement of this angle would have
consequences for this case of texture 5 zero mass matrices for
Dirac neutrinos.

Considering the texture 5 zero $D_{\nu}=0$ case the possibility of
$M_l$ having Fritzsch-like structure reveals several interesting
facts, as can be seen from Table (\ref{tab1dir}). A comparison
with the earlier cases shows both the lower and upper limits of
$m_{\nu_1}$ have higher values. Interestingly, for this case the
lower limit of $s_{13}$ becomes almost 0, implying that
measurement of this angle can lead to interesting implications for
the texture specific mass matrices considered here. Also, it may
be noted that out of all the three cases, this case has the widest
range of the Dirac-like CP violating phase $\delta_l$. The PMNS
matrix corresponding to this case is quite similar to the one
presented in equation (\ref{dirmat}), except for somewhat wider
ranges of the elements $U_{e 3}$, $U_{\mu 2}$, $U_{\tau 1}$ and
$U_{\tau 2}$. This can be understood by noting that $D_l$ can take
much wider variation compared to $D_{\nu}$.

It may be added that in case one carries out an exercise regarding
the variation of phases $\phi_1$ and $\phi_2$ w.r.t the mixing
angle $s_{13}$, we find results similar to the ones obtained for
the texture 6 zero case, hence we have not presented the same.

\section{Summary and conclusions\label{summ}}
To summarize, for Dirac neutrinos, using Fritzsch-like texture 6
zero and 5 zero mass matrices, detailed predictions for 15
distinct possible cases pertaining to normal/inverted hierarchy as
well as degenerate scenario of neutrino masses have been carried
out. Interestingly, all the presently considered cases pertaining
to inverted hierarchy and degenerate scenario seem to be ruled
out.

In the normal hierarchy cases, when the charged lepton mass matrix
$M_l$ is assumed to be in flavor basis, the texture 6 zero and the
texture 5 zero $D_{\nu}=0$ case are again ruled out. For the
viable texture 6 zero and 5 zero cases, we find that the ranges of
the neutrino masses $m_{\nu_1}$, $m_{\nu_2}$, $m_{\nu_3}$ as well
as of the mixing angle $s_{13}$ would have implications for the
texture specific cases considered here. Interestingly, the lower
limit of $s_{13}$ for the texture 5 zero $D_{\nu}=0$ case shows an
appreciable difference as compared to the lower limits of $s_{13}$
for the texture 6 zero and texture 5 zero $D_l=0$ cases.
Similarly, the Dirac-like CP violating phase $\delta_l$ shows
interesting behaviour, e.g., as compared to the texture 6 zero
case, the texture 5 zero cases allow comparatively a larger range
of $\delta_l$, being widest in the texture 5 zero $D_{\nu}=0$
case. The restricted range of $\delta_l$, in spite of full
variation to phases $\phi_1$ and $\phi_2$, seems to be due to
texture structure, hence, any information about $\delta_l$ would
have important implications.

  \vskip 0.5cm
{\bf Acknowledgements} \\ The authors would like to thank S.D.
Sharma and Sanjeev Kumar for useful discussions. GA and MG would
like to thank DST, Government of India and DAE, BRNS for financial
support respectively. GA would also like to thank the Chairman,
Department of Physics for providing facilities to work in the
department. MR and RV would like to thank the Director, UIET and
the Director, RIEIT respectively, for providing facilities to
work.

\vskip 1.5cm

\onehalfspacing {\topskip 2cm \bc {\large\bf{Appendix}} \ec
\renewcommand\theequation{A.\arabic{equation}}
  \setcounter{equation}{0}

In this Appendix, we present the elements of the
Pontecorvo-Maki-Nakagawa-Sakata (PMNS) mixing matrix in the case
of Dirac neutrinos corresponding to texture 4 zero mass matrices.
The corresponding relations for the texture 5 and 6 zero mass
matrices can be easily derived from these. For example,
considering both $D_l$ and $D_{\nu}$ to be zero the relations for
texture 6 zero mass matrices are obtained, whereas for the texture
5 zero mass matrices either $D_l$ or $D_{\nu}$ is considered to be
zero. The expressions for the elements of the PMNS mixing matrix
are given by
 \beqn
 U_{e1}=\sqrt{\frac{m_1 (-D_{\nu} - m_2 +
m_3)}{(m_1+m_2)(-m_1+m_3)}} \sqrt{\frac{m_e (-D_{l} - m_{\mu} +
m_{\tau})}{(m_e+m_{\mu})(-m_e+m_{\tau})}}+~~~~~~~~~~~ \nonum \\
\sqrt{\frac{m_2 m_3 (-D_{\nu} - m_2 +
m_3)}{C_{\nu}(m_1+m_2)(-m_1+m_3)}} \sqrt{\frac{m_{\mu}
m_{\tau}(-D_{l} - m_{\mu} +
m_{\tau})}{C_l(m_e+m_{\mu})(-m_e+m_{\tau})}}~e^{i \phi_1} + \nonum
\\ \sqrt{\frac{m_1 (D_{\nu} - m_1 + m_2)(-D_{\nu} + m_1 +
m_3)}{C_{\nu}(m_1+m_2)(-m_1+m_3)}} \times
~~~~~~~~~~~~~~~~~~~~~~~~~~\nonum \\ \sqrt{\frac{m_e (D_{l} - m_e +
m_{\mu})(-D_{l} + m_e +
m_{\tau})}{C_{l}(m_e+m_{\mu})(-m_e+m_{\tau})}}~e^{i
\phi_2}~~~~~~~~~~~~~~~~~~~~~~~~~~~
 \\
 \nonum \\
  \nonum \\
U_{e2}=\sqrt{\frac{m_2 (-D_{\nu} + m_1 +
m_3)}{(m_1+m_2)(m_2+m_3)}} \sqrt{\frac{m_e (-D_{l} - m_{\mu} +
m_{\tau})}{(m_e+m_{\mu})(-m_e+m_{\tau})}}-~~~~~~~~~~~ \nonum \\
\sqrt{\frac{m_1 m_3 (-D_{\nu} + m_1 +
m_3)}{C_{\nu}(m_1+m_2)(m_2+m_3)}} \sqrt{\frac{m_{\mu}
m_{\tau}(-D_{l} - m_{\mu} +
m_{\tau})}{C_l(m_e+m_{\mu})(-m_e+m_{\tau})}}~e^{i \phi_1} + \nonum
\\ \sqrt{\frac{m_2 (D_{\nu} - m_1 + m_2)(-D_{\nu} - m_2 +
m_3)}{C_{\nu}(m_1+m_2)(m_2+m_3)}} \times
~~~~~~~~~~~~~~~~~~~~~~~~~~\nonum \\ \sqrt{\frac{m_e (D_{l} - m_e +
m_{\mu})(-D_{l} + m_e +
m_{\tau})}{C_{l}(m_e+m_{\mu})(-m_e+m_{\tau})}}~e^{i
\phi_2}~~~~~~~~~~~~~~~~~~~~~~~~~~~
 \\
  \nonum \\
  \nonum \\
  U_{e3}=\sqrt{\frac{m_3 (D_{\nu} - m_1 +
m_2)}{(-m_1+m_3)(m_2+m_3)}} \sqrt{\frac{m_e (-D_{l} - m_{\mu} +
m_{\tau})}{(m_e+m_{\mu})(-m_e+m_{\tau})}}+~~~~~~~~~~~ \nonum \\
\sqrt{\frac{m_1 m_2 (D_{\nu} - m_1 +
m_2)}{C_{\nu}(-m_1+m_3)(m_2+m_3)}} \sqrt{\frac{m_{\mu}
m_{\tau}(-D_{l} - m_{\mu} +
m_{\tau})}{C_l(m_e+m_{\mu})(-m_e+m_{\tau})}}~e^{i \phi_1} - \nonum
\\ \sqrt{\frac{m_3 (-D_{\nu} + m_1 + m_3)(-D_{\nu} - m_2 +
m_3)}{C_{\nu}(-m_1+m_3)(m_2+m_3)}} \times
~~~~~~~~~~~~~~~~~~~~~~~~~~\nonum \\ \sqrt{\frac{m_e (D_{l} - m_e +
m_{\mu})(-D_{l} + m_e +
m_{\tau})}{C_{l}(m_e+m_{\mu})(-m_e+m_{\tau})}}~e^{i
\phi_2}~~~~~~~~~~~~~~~~~~~~~~~~~~~
 \\
  \nonum \\
  \nonum \\
 U_{\mu 1}=\sqrt{\frac{m_1 (-D_{\nu} - m_2 +
m_3)}{(m_1+m_2)(-m_1+m_3)}} \sqrt{\frac{m_{\mu} (-D_{l} + m_{e} +
m_{\tau})}{(m_e+m_{\mu})(m_{\mu}+m_{\tau})}}-~~~~~~~~~~~ \nonum
\\ \sqrt{\frac{m_2 m_3 (-D_{\nu} - m_2 +
m_3)}{C_{\nu}(m_1+m_2)(-m_1+m_3)}} \sqrt{\frac{m_{e}
m_{\tau}(-D_{l} + m_{e} +
m_{\tau})}{C_l(m_e+m_{\mu})(m_{\mu}+m_{\tau})}}~e^{i \phi_1} +
\nonum
\\ \sqrt{\frac{m_1 (D_{\nu} - m_1 + m_2)(-D_{\nu} + m_1 +
m_3)}{C_{\nu}(m_1+m_2)(-m_1+m_3)}} \times
~~~~~~~~~~~~~~~~~~~~~~~~~~\nonum \\ \sqrt{\frac{m_{\mu} (D_{l} -
m_e + m_{\mu})(-D_{l} - m_{\mu} +
m_{\tau})}{C_{l}(m_e+m_{\mu})(m_{\mu}+m_{\tau})}}~e^{i
\phi_2}~~~~~~~~~~~~~~~~~~~~~~~~~~~
 \\
 \nonum \\
  \nonum \\
U_{\mu 2}=\sqrt{\frac{m_2 (-D_{\nu} + m_1 +
m_3)}{(m_1+m_2)(m_2+m_3)}} \sqrt{\frac{m_{\mu} (-D_{l} + m_{e} +
m_{\tau})}{(m_e+m_{\mu})(m_{\mu}+m_{\tau})}}+~~~~~~~~~~~ \nonum
\\ \sqrt{\frac{m_1 m_3 (-D_{\nu} + m_1 +
m_3)}{C_{\nu}(m_1+m_2)(m_2+m_3)}} \sqrt{\frac{m_{e}
m_{\tau}(-D_{l} + m_{e} +
m_{\tau})}{C_l(m_e+m_{\mu})(m_{\mu}+m_{\tau})}}~e^{i \phi_1} +
\nonum
\\ \sqrt{\frac{m_2 (D_{\nu} - m_1 + m_2)(-D_{\nu} - m_2 +
m_3)}{C_{\nu}(m_1+m_2)(m_2+m_3)}} \times
~~~~~~~~~~~~~~~~~~~~~~~~~~\nonum \\ \sqrt{\frac{m_{\mu} (D_{l} -
m_e + m_{\mu})(-D_{l} - m_{\mu} +
m_{\tau})}{C_{l}(m_e+m_{\mu})(m_{\mu}+m_{\tau})}}~e^{i
\phi_2}~~~~~~~~~~~~~~~~~~~~~~~~~~~
 \\
 \nonum \\
  \nonum \\
U_{\mu 3}=\sqrt{\frac{m_3 (D_{\nu} - m_1 +
m_2)}{(-m_1+m_3)(m_2+m_3)}} \sqrt{\frac{m_{\mu} (-D_{l} + m_{e} +
m_{\tau})}{(m_e+m_{\mu})(m_{\mu}+m_{\tau})}}-~~~~~~~~~~~ \nonum
\\ \sqrt{\frac{m_1 m_2 (D_{\nu} - m_1 +
m_2)}{C_{\nu}(-m_1+m_3)(m_2+m_3)}} \sqrt{\frac{m_{e}
m_{\tau}(-D_{l} + m_{e} +
m_{\tau})}{C_l(m_e+m_{\mu})(m_{\mu}+m_{\tau})}}~e^{i \phi_1} -
\nonum
\\ \sqrt{\frac{m_3 (-D_{\nu} + m_1 + m_3)(-D_{\nu} - m_2 +
m_3)}{C_{\nu}(-m_1+m_3)(m_2+m_3)}} \times
~~~~~~~~~~~~~~~~~~~~~~~~~~\nonum \\ \sqrt{\frac{m_{\mu} (D_{l} -
m_e + m_{\mu})(-D_{l} - m_{\mu} +
m_{\tau})}{C_{l}(m_e+m_{\mu})(m_{\mu}+m_{\tau})}}~e^{i
\phi_2}~~~~~~~~~~~~~~~~~~~~~~~~~~~
 \\
 \nonum \\
  \nonum \\
  U_{\tau 1}=\sqrt{\frac{m_1 (-D_{\nu} - m_2 +
m_3)}{(m_1+m_2)(-m_1+m_3)}} \sqrt{\frac{m_{\tau} (D_{l} - m_{e} +
m_{\mu})}{(-m_e-m_{\tau})(m_{\mu}+m_{\tau})}}+~~~~~~~~~~~ \nonum
\\ \sqrt{\frac{m_2 m_3 (-D_{\nu} - m_2 +
m_3)}{C_{\nu}(m_1+m_2)(-m_1+m_3)}} \sqrt{\frac{m_{e} m_{\mu}(D_{l}
- m_{e} + m_{\mu})}{C_l(-m_e+m_{\tau})(m_{\mu}+m_{\tau})}}~e^{i
\phi_1} - \nonum
\\ \sqrt{\frac{m_1 (D_{\nu} - m_1 + m_2)(-D_{\nu} + m_1 +
m_3)}{C_{\nu}(m_1+m_2)(-m_1+m_3)}} \times
~~~~~~~~~~~~~~~~~~~~~~~~~~\nonum \\ \sqrt{\frac{m_{\tau} (-D_{l} +
m_e + m_{\tau})(-D_{l} - m_{\mu} +
m_{\tau})}{C_{l}(-m_e+m_{\tau})(m_{\mu}+m_{\tau})}}~e^{i
\phi_2}~~~~~~~~~~~~~~~~~~~~~~~~~~~
 \\
 \nonum \\
  \nonum \\
  U_{\tau 2}=\sqrt{\frac{m_2 (-D_{\nu} + m_1 +
m_3)}{(m_1+m_2)(m_2+m_3)}} \sqrt{\frac{m_{\tau} (D_{l} - m_{e} +
m_{\mu})}{(-m_e+m_{\tau})(m_{\mu}+m_{\tau})}}-~~~~~~~~~~~ \nonum
\\ \sqrt{\frac{m_1 m_3 (-D_{\nu} + m_1 +
m_3)}{C_{\nu}(m_1+m_2)(m_2+m_3)}} \sqrt{\frac{m_{e} m_{\mu}(D_{l}
- m_{e} + m_{\mu})}{C_l(-m_e+m_{\tau})(m_{\mu}+m_{\tau})}}~e^{i
\phi_1} - \nonum
\\ \sqrt{\frac{m_2 (D_{\nu} - m_1 + m_2)(-D_{\nu} - m_2 +
m_3)}{C_{\nu}(m_1+m_2)(m_2+m_3)}} \times
~~~~~~~~~~~~~~~~~~~~~~~~~~\nonum \\ \sqrt{\frac{m_{\tau} (-D_{l} +
m_e + m_{\tau})(-D_{l} - m_{\mu} +
m_{\tau})}{C_{l}(-m_e+m_{\tau})(m_{\mu}+m_{\tau})}}~e^{i
\phi_2}~~~~~~~~~~~~~~~~~~~~~~~~~~~
 \\
 \nonum \\
  \nonum \\
  U_{\tau 3}=\sqrt{\frac{m_3 (D_{\nu} - m_1 +
m_2)}{(-m_1+m_3)(m_2+m_3)}} \sqrt{\frac{m_{\tau} (D_{l} - m_{e} +
m_{\mu})}{(-m_e+m_{\tau})(m_{\mu}+m_{\tau})}}+~~~~~~~~~~~ \nonum
\\ \sqrt{\frac{m_1 m_2 (D_{\nu} - m_1 +
m_2)}{C_{\nu}(-m_1+m_3)(m_2+m_3)}} \sqrt{\frac{m_{e} m_{\mu}(D_{l}
- m_{e} + m_{\mu})}{C_l(-m_e+m_{\tau})(m_{\mu}+m_{\tau})}}~e^{i
\phi_1} - \nonum
\\ \sqrt{\frac{m_3 (-D_{\nu} + m_1 + m_3)(-D_{\nu} - m_2 +
m_3)}{C_{\nu}(-m_1+m_3)(m_2+m_3)}} \times
~~~~~~~~~~~~~~~~~~~~~~~~~~\nonum \\ \sqrt{\frac{m_{\tau} (-D_{l} +
m_e + m_{\tau})(-D_{l} - m_{\mu} +
m_{\tau})}{C_{l}(-m_e+m_{\tau})(m_{\mu}+m_{\tau})}}~e^{i
\phi_2}~~~~~~~~~~~~~~~~~~~~~~~~~~~ \eeqn

\newpage

\begin{table}
\bc {\renewcommand{\arraystretch}{1.4}
\begin{tabular}{|c|c|c|c|} \hline
 & 6 zero & 5 zero $D_l=0$ & 5 zero $D_{\nu}=0$\\

 & & ($M_l$ 3 zero, $M_{\nu D}$ 2 zero) &  ($M_l$ 2 zero, $M_{\nu D}$ 3
 zero)
 \\ \hline

$m_{\nu_1}$ & $\sim$ 0.0025 & 0.00032 - 0.0063  & 0.0025 - 0.0079
\\
 $m_{\nu_2}$ & 0.0093  - 0.0096 & 0.0086 - 0.0112  & 0.0089 -
0.0122
\\
$m_{\nu_3}$ &$\sim$ 0.0423  & 0.0421 - 0.0550 & 0.0422 - 0.0552
\\
$s_{13}$ & 0.007 - 0.026  & 0.005 - 0.160 &0.0001 - 0.135\\

 $J_l$ &$\sim$ 0 - 0.005  & $\sim$ 0 - 0.027 & $\sim$ 0 - 0.028 \\

$\delta_l$ & $3.6^{\circ}$ - 69.2$^{\circ}$  &$0^{\circ}$ -
80.2$^{\circ}$
 &$0^{\circ}$ - 90.0$^{\circ}$

\\ \hline

\end{tabular}} \ec
\caption{Calculated ranges for neutrino mass and mixing parameters
obtained by varying $\phi_1$ and $\phi_2$ from 0 to 2$\pi$ for the
normal hierarchy cases of Dirac neutrinos. Inputs have been
defined in the text. All masses are in $\rm{eV}$. }
\label{tab1dir}
\end{table}

\begin{figure}[hbt]
\centerline{\epsfysize=4.in\epsffile{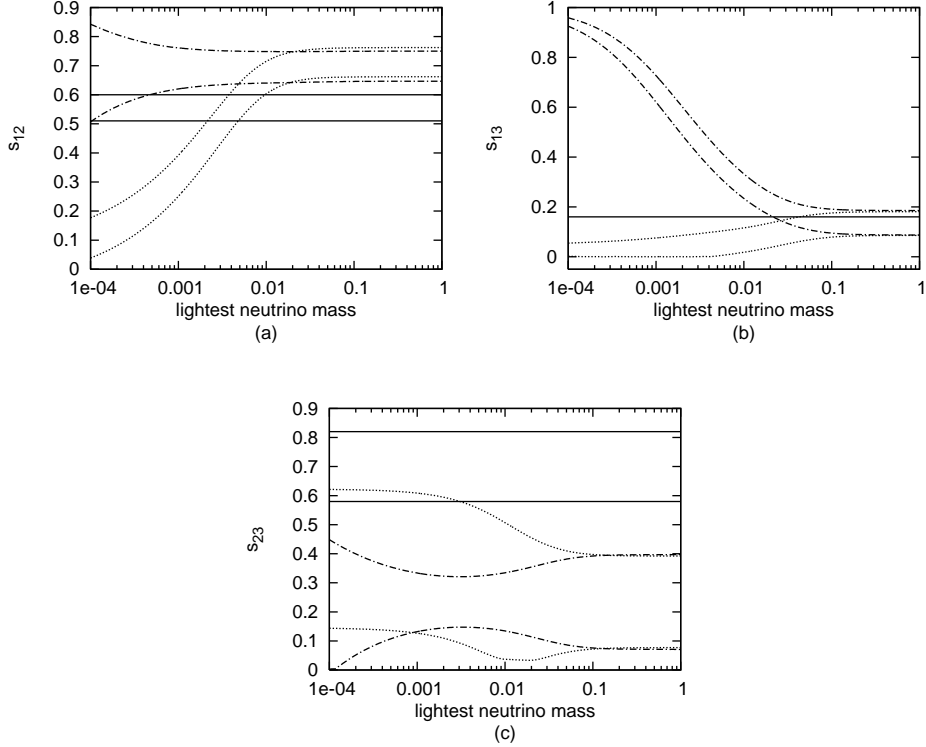}}
 \caption{Plots showing variation of mixing angles
$s_{12}$, $s_{13}$ and $s_{23}$ with lightest neutrino mass for
texture 6 zero case of Dirac neutrinos. The dotted lines and the
dot-dashed lines depict the limits obtained assuming normal and
inverted hierarchy respectively, the solid horizontal lines show
the 3$\sigma$ limits of $s_{23}$ given in equation (\ref{s13}).}
\label{nhih6zdir}
\end{figure}

\begin{figure}[hbt]
\centerline{\epsfysize=4.in\epsffile{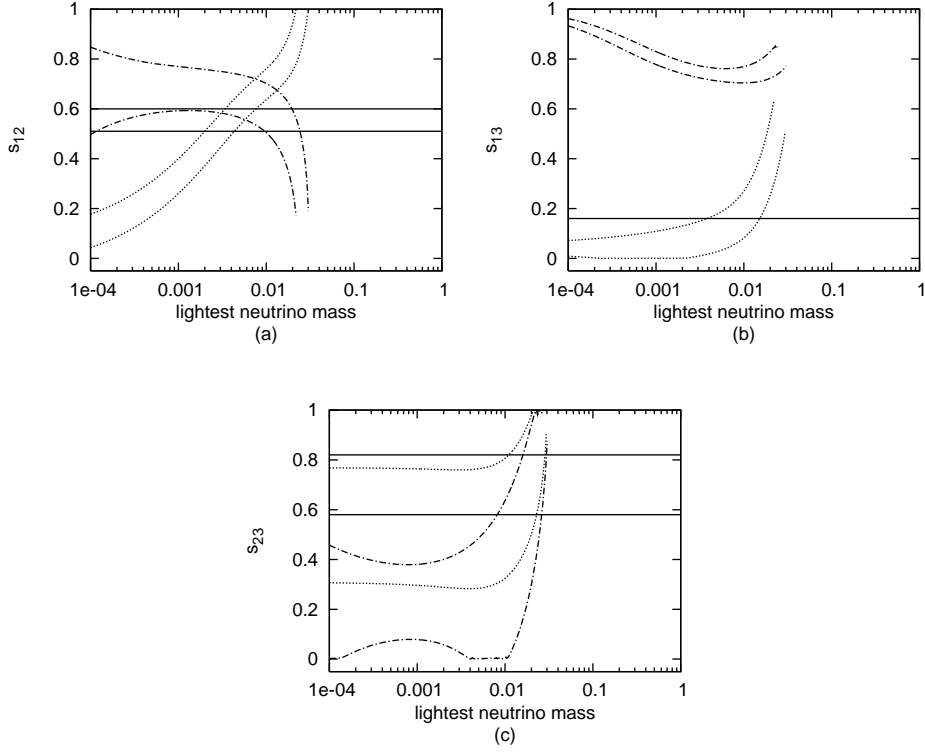}}
 \caption{Plots showing variation of mixing angles
$s_{12}$, $s_{13}$ and $s_{23}$ with lightest neutrino mass for
texture 5 zero $D_l=0$ case of Dirac neutrinos, with a value
$D_{\nu}= \sqrt{m_{\nu_3}}$. The representations of the curves
remain the same as in Figure (\ref{nhih6zdir}).}
\label{nhih5zdndir}
\end{figure}

\begin{figure}[hbt]
\centerline{\epsfysize=2.2 in\epsffile{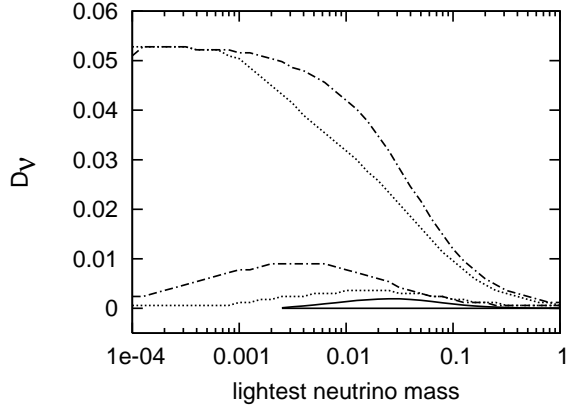}}
 \caption{Plots showing allowed parameter space for the
  three mixing angles in the
$D_{\nu}-$lightest neutrino mass plane, for texture 5 zero $D_l=0$
case of Dirac neutrinos for the inverted hierarchy, with $D_{\nu}$
being varied from 0 to a value such that
$D_{\nu}<\sqrt{m_{\nu_1}}-\sqrt{m_{\nu_3}}$.
 Dotted lines
depict allowed parameter space for $s_{12}$, dot-dashed lines
depict allowed parameter space for $s_{23}$ and solid lines depict
allowed parameter space for $s_{13}$.} \label{dmlimitsihdir}
\end{figure}

\begin{figure}[hbt]
\centerline{\epsfysize=4.in\epsffile{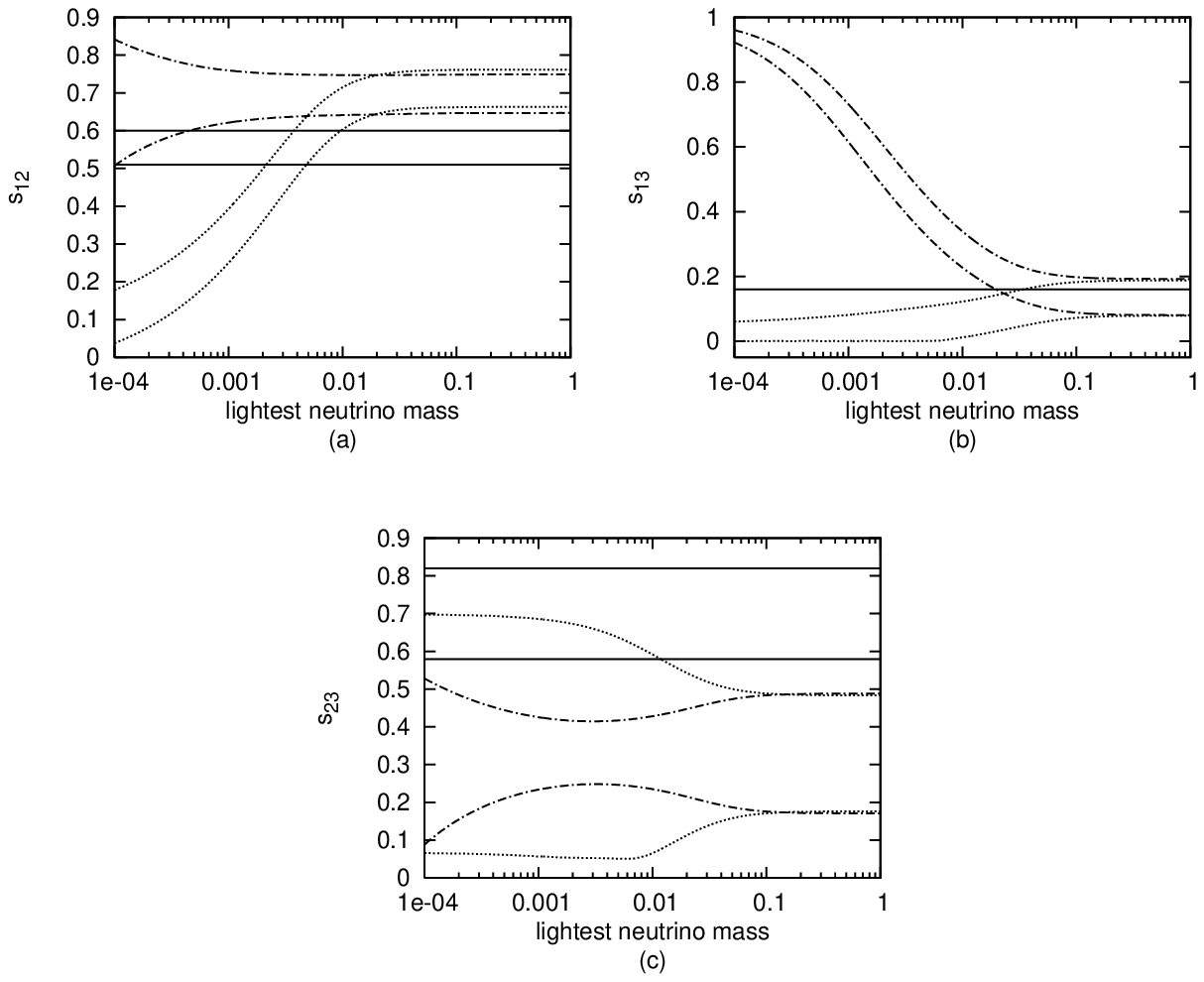}}
 \caption{Plots showing variation of mixing angles
$s_{12}$, $s_{13}$ and $s_{23}$ with lightest neutrino mass for
texture 5 zero $D_{\nu}=0$ case of Dirac neutrinos. The
representations of the curves remain the same as in Figure
(\ref{nhih6zdir}).} \label{nhih5zdldir}
\end{figure}

\begin{figure}[hbt]
\centerline{\epsfysize=3.2 in\epsffile{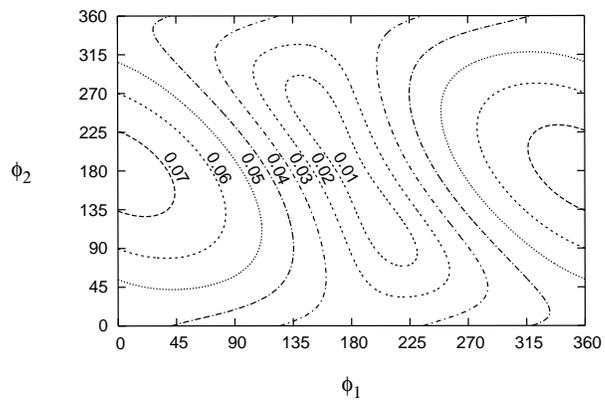}}
 \caption{The contours of $s_{13}$ in $\phi_1 - \phi_2$ plane for
texture 6 zero matrices for the normal hierarchy case of Dirac
neutrinos.} \label{s13cont6zdir}
\end{figure}

\end{document}